\documentclass[11pt]{article}
\usepackage{times}
\usepackage{geometry}
\usepackage[utf8]{inputenc}
\usepackage{enumitem,amssymb}
\usepackage{ragged2e}
\newlist{thematic}{itemize}{8}
\setlist[thematic]{label=$\square$}
\usepackage{pifont}

\newcommand{\myvspace}{\vspace{0.1cm}}
\usepackage{enumitem}
\setlist[itemize]{leftmargin=*}
\usepackage{fancyhdr}
\usepackage[USenglish]{babel}
\usepackage[utf8]{inputenc}
\usepackage[T1]{fontenc}
\usepackage{epsfig}
\usepackage{wrapfig}
\usepackage{color}
\usepackage[vflt]{floatflt}
\usepackage{longtable}
\usepackage{caption}
\usepackage{jheppub}   
\usepackage[dvipsnames,svgnames,x11names]{xcolor}
\usepackage[all]{hypcap} 
\usepackage{amssymb,amsmath}
\usepackage{longtable}
\usepackage{amssymb}
\usepackage{amsmath, xspace}
\usepackage{graphics,graphicx} 
\usepackage{rotating}
\usepackage{color}
\usepackage{xcolor}
\usepackage{multirow}
\usepackage{subfigure}
\usepackage{sectsty}
\usepackage[outercaption]{sidecap}
\sidecaptionvpos{figure}{c}

\usepackage{tikz}

\usepackage{xcolor}
\definecolor{cobalt}{rgb}{0., 0.35, 0.56}
\usepackage{hyperref}

\usepackage{enumitem}
\setlist[enumerate]{itemsep=0pt, parsep=0pt}

\usepackage{natbib}

\definecolor{DarkGreen}{rgb}{0.0, 0.3, 0.0}
\definecolor{purple}{rgb}{0.5, 0.0, 0.5}
\definecolor{red}{rgb}{1, 0.0, 0.0}
\definecolor{green}{rgb}{0, 1.0, 0.0}

\def\3he{$^3{\rm He}$}

\def\lsim{\mathrel{\lower2.5pt\vbox{\lineskip=0pt\baselineskip=0pt
           \hbox{$<$}\hbox{$\sim$}}}}

\def\gsim{\mathrel{\lower2.5pt\vbox{\lineskip=0pt\baselineskip=0pt
           \hbox{$>$}\hbox{$\sim$}}}}

\begin{document}
\raggedright
\huge

Bridging the UV Gap

The HST Ultraviolet Foundation for Star Formation Science in the Era of Roman, Euclid, and HWO

\bigskip
\normalsize

\raggedright

\myvspace

\textbf{Authors:} 
Fatemeh Zahra Majidi (fatemeh.majidi@inaf.it, INAF-OACN, Italy); Amelia Bayo (ESO-Garching, Germany); Katia Biazzo (INAF-Osservatorio Astronomico di Roma, Italy);

\textbf{Endorsers:}  J. M. Alcal\'a (INAF-OACN, Italy), K. France (University of Colorado Boulder, USA), E. Gaidos (UH IfA, USA), M. G. Guarcello (INAF-OAPA, Italy).

\myvspace

\begin{figure}[!ht]
    \centering
    \includegraphics[width=0.9\linewidth]{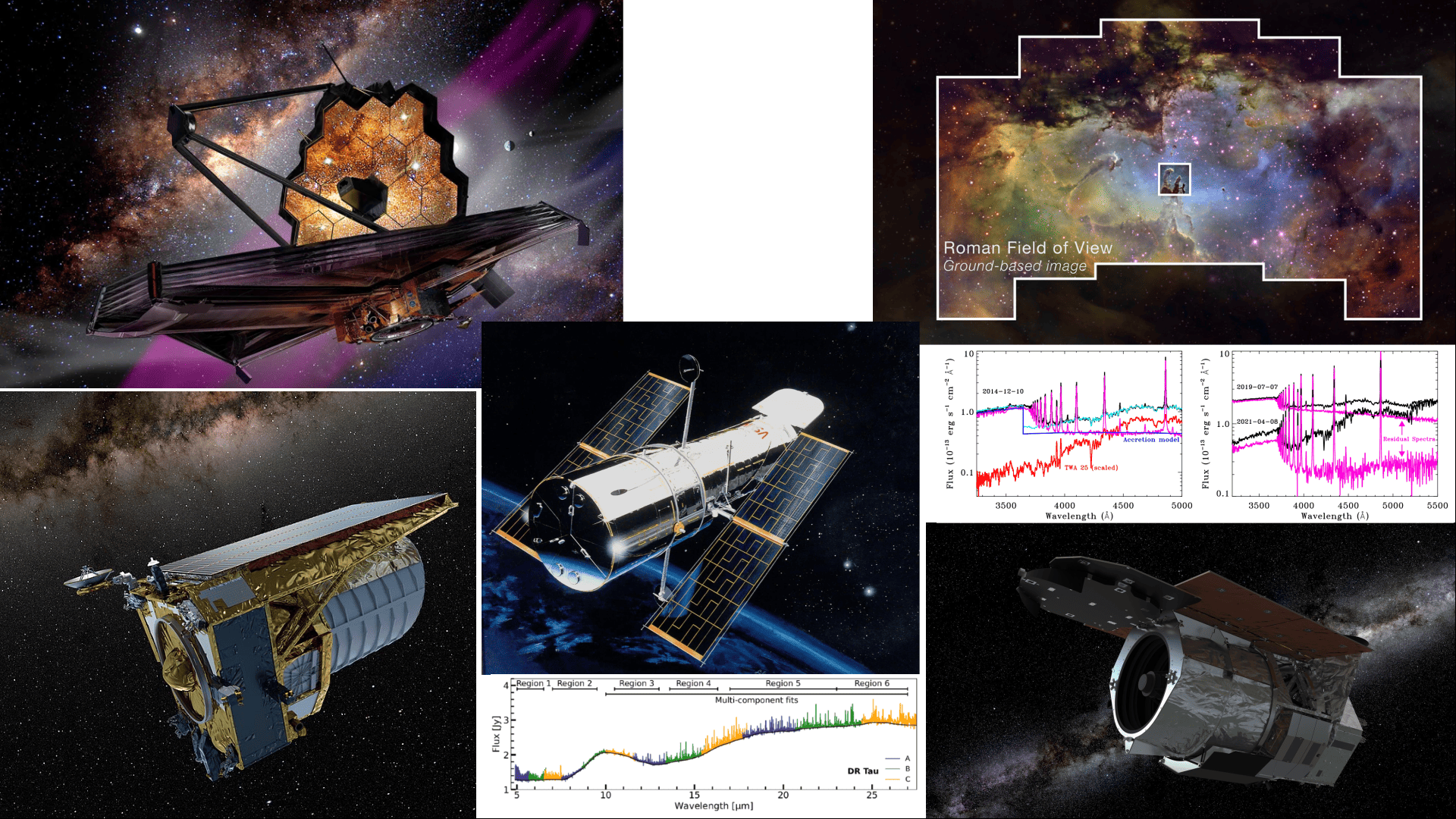}
    \caption*{\small \textbf{Figure.} JWST (Scientific American), Roman (Caltech/IPAC), Euclid (ESA), HST, and Roman’s Field of View vs. HST’s (NASA). Spectra of TW Hya (Right, \cite{Herczeg_2023}) and DR Tau (Bottom, \cite{2024A&A...689A.330T}). }
    \label{fig:placeholder}
\end{figure}

\textbf{Abstract:} As we enter the 2030s, the astronomical landscape will be dominated by large-scale infrared (IR) and optical surveys led by JWST, Euclid, and the Nancy Grace Roman Space Telescope. While these facilities provide unprecedented views of the dusty environments of nearby star-forming regions, they are fundamentally limited in their ability to probe the high-energy physics of accretion, magnetospheric activity, and disk photoevaporation. 
This white paper argues for the critical continued use of the Hubble Space Telescope (HST) Space Telescope Imaging Spectrograph (STIS) and Cosmic Origins Spectrograph (COS) to bridge the "UV Gap." We demonstrate that UV spectroscopy is the only direct method for characterizing the feedback mechanisms that determine planet habitability and stellar maturation, serving as a mandatory scientific bridge toward the Habitable Worlds Observatory (HWO). 
The study of star formation stands at a critical intersection of multiple scientific disciplines, linking the high-energy physics of stellar birth to the chemical evolution of protoplanetary disks and the eventual habitability of exoplanets. As such, it represents one of the most compelling and essential science cases for the continued allocation of HST resources. Ensuring that HST provides high-resolution UV spectroscopic data now is a fundamental requirement for the success of future flagship missions, as these data provide the unique physical context that infrared observations alone cannot achieve.

\section{The Imminent $''$UV Gap$''$}

The upcoming decade heralds a paradigm shift in observational astronomy, characterized by the high-cadence, deep-field infrared (IR) capabilities of JWST, Euclid, and the Nancy Grace Roman Space Telescope. While these missions will catalog young stellar objects (YSOs) and brown dwarfs with unprecedented statistical significance, the community faces an impending "UV Gap" that threatens to decouple these photometric detections from their underlying high-energy physics. It’s noteworthy that the bluest wavelength reachable from the ground will be provided through ESO’s CUBES (~300 nm, \cite{2023MmSAI..94b.281C}), but accessing the UV regime from the ground isn’t an optimal approach because the Earth's atmosphere absorbs almost all incoming cosmic UV radiation.

\noindent As the HST enters its late-mission phase, the transition to an IR-dominated landscape without a concurrent UV spectroscopic facility creates a critical diagnostic deficit.
While several small-scale missions and CubeSats have emerged to address the need for ultraviolet observations, the limited spatial and spectral resolution offered by these missions cannot contribute significantly to the science cases described below and heavily supported by the HST. These miniature observatories serve as vital technology demonstrators and are capable of monitoring the brightest stellar targets; however, they lack the collecting area and aperture required to reach the magnitudes and depths necessary for characterizing distant star-forming regions. Whereas CubeSats are limited to the most luminous local stars, the high-sensitivity optics of HST/STIS and COS allow for the spectroscopic dissection of faint, young stellar objects and substellar members across entire molecular clouds (\cite{2012ApJ...756...24S}; \cite{Muzic_2014}) —populations that will form the core of the Roman and Euclid catalogs (\href{https://roman.gsfc.nasa.gov/science/gps/Roman%20Galactic%20Plane%20Survey%20Definition%20Committee%20Report%20(submitted).pdf}{Roman Galactic Plane Survey Definition Committee Report}; \cite{2025A&A...697A...7M}). Relying solely on limited-capacity small-sats would effectively blind the community to the lower-mass and more distant regimes of star formation, leaving a significant portion of the infrared-detected census without the high-resolution UV context required for a complete physical interpretation.

\noindent The fundamental drivers of stellar and planetary maturation (specifically the FUV/NUV flux) dictate gas-phase chemistry and the thermal evolution of protoplanetary disks (\cite{2012ApJ...744...60G}; \cite{2023ASPC..534..567P}). This paper establishes that bridging the UV Gap is a scientific prerequisite for synthesizing the multi-wavelength data of the 2030s. We demonstrate how HST/COS and STIS provide one of the most direct ways to infer mass accretion rates ($\dot{M}_{acc}$, \cite{Espaillat_2022}) and the 105 K plasma temperatures (\cite{Ardila_2013}) within magnetospheric funnels, offering the essential calibration required to interpret the millions of objects discovered by Roman and Euclid (to be operational well into the 2030s).

\noindent Furthermore, we explore the mechanical synergy between JWST’s probes of the disk mid-plane and HST’s unique ability to resolve the FUV-irradiated skin of the protoplanetary disk, where photoevaporative winds determine the inventory of volatiles available for planetesimal growth (\cite{2017RSOS....470114E}). Extending this analysis to the substellar regime, we utilize UV spectroscopic diagnostics to disentangle the formation pathways of free-floating planets and brown dwarfs (\cite{Zhang_2025}) – expected from Euclid's and Roman’s wide surveys (\cite{2020AJ....160..123J}; \cite{2022A&A...664A.136B}). Ultimately, this framework positions current HST legacy programs as a mandatory temporal anchor for the HWO, providing the decades-long UV baseline necessary to contextualize the atmospheric stability and prebiotic potential of the planetary systems HWO will eventually target.

\section{Accretion and Magnetospheric Physics in the UV Regime}

In the standard paradigm of magnetospheric accretion (assumed to be applicable to solar and low-mass young stars), the inner protoplanetary disk is truncated by the stellar magnetosphere at a radius close to the corotation radius, where the Keplerian angular velocity of the disk matches the stellar rotation rate. Gas from the inner disk is subsequently funneled along magnetic field lines toward the stellar surface, undergoing near free-fall before impacting the photosphere at supersonic velocities (\cite{1998apsf.book.....H}).

\subsection{The High-Energy Shock Interface}

The resulting accretion shock produces localized hot spots with characteristic temperatures reaching $\sim10^5$–$10^6$ K (\cite{Ardila_2013}), emitting strongly in the FUV and NUV continuum. Optical diagnostics such as $H_{\alpha}$ and the Ca II infrared triplet are commonly employed as accretion tracers (\cite{2014A&A...561A...2A, 2017A&A...600A..20A}; \cite{2017A&A...604A.127M}); however, these lines originate in spatially extended and lower-density regions that are significantly affected by stellar winds, outflows, and chromospheric activity (\cite{2023ASPC..534..539M}; \cite{2023ASPC..534..567P}). In contrast, the HST instruments COS and STIS provide access to resonance transitions of highly ionized species, particularly C IV 1548–1550 \AA \hspace{0.02cm} and Si IV 1394–1403 \AA, which directly probe the hot post-shock plasma near the base of the accretion funnel flows (\cite{Carvalho_2024}). 
In the context of synergic observations with Roman and Euclid, HST’s observations is of great importance to produce multiband data. According to \cite{2026arXiv260519710D}, throughout an accretion outburst, the redest optical and near-infrared lightcurves show sensitivity to heating in the accretion shocks and inner gas disk, while mid-infrared lightcurves are more responsive to the location and heating of the innermost dust disk. Multiband data in this context will provide valuable insight into star-magnetosphere-disk interactions throughout the outburst cycle. 

\subsection{Model-Independent Accretion Rates}

Direct measurement of the UV excess continuum (“veiling”, \cite{2021A&A...652A..72A}), which partially fills in or obscures photospheric absorption features, provides the most direct observational constraint on the accretion luminosity, $L_{acc}$ (\cite{2011ApJ...743..105I,Ingleby_2013}). While deriving $\dot{M}_{acc}$ still requires assumptions regarding stellar parameters and accretion geometry, UV continuum measurements obtained with HST constitute the fundamental calibration standard for empirical accretion diagnostics. In the absence of these UV-based calibrations, large photometric surveys conducted by Roman and Euclid will necessarily rely on indirect empirical relations that can introduce uncertainties approaching an order of magnitude, particularly across heterogeneous star-forming environments. Establishing a robust “UV gold standard” is therefore essential for accurately characterizing the evolution of $\dot{M}_{acc}$ in millions of YSOs throughout the Galaxy. 

\section{Dissecting the $''$Top-Down$''$ Evolution of Protoplanetary Disks}

The formation of planetary systems is fundamentally governed by the competition between gas accretion onto forming planets and the dispersal of the protoplanetary disk itself. While James Webb Space Telescope provides unprecedented sensitivity to dust emission and cool molecular gas within the disk mid-plane, the ultimate dispersal of the disk is thought to be regulated largely by high-energy irradiation and photoevaporative mass loss from the disk surface layers (\cite{2014prpl.conf..475A}). EUV, FUV, and X-ray radiation from the central young star heat the upper layers of the disk atmosphere, driving thermally launched photoevaporative winds that can disperse the gaseous component of the disk on characteristic timescales of a few to $\sim$10 Myr, consistent with observed protoplanetary disk lifetimes (\cite{2001ApJ...553L.153H}, \cite{2018MNRAS.477.5191R}). FUV irradiation is particularly important for heating molecular surface layers, whereas EUV photons primarily ionize hydrogen in the inner wind regions (\cite{Gorti_2009}).

\noindent The HST instrument STIS is uniquely capable of resolving fluorescent H2 emission and FUV CO band emission, which trace the warm molecular gas in the irradiated surface layers of the disk (\cite{2012arXiv1208.2270F}). These diagnostics provide direct constraints on the kinematics, excitation conditions, and mass-loss processes operating in the disk atmosphere. By combining HST observations of the evolving gaseous component with JWST measurements of dust growth, settling, and mineralogy in the disk mid-plane, it becomes possible to construct a more complete multidimensional picture of the disk’s thermal, dynamical, and chemical structure (\cite{2024ApJ...963..158P}; \cite{2024PASP..136e4302H}). Such constraints are critical for understanding the evolution of the disk gas-to-dust ratio, the volatile inventories incorporated into giant planets, and the efficiency of water and organic delivery to terrestrial planets (\cite{OBERG20211})—the same class of potentially habitable worlds that future observatories such as HWO are designed to characterize.

\noindent Both science cases outlined in Sections 2 and 3 have been substantially advanced by the ODYSSEUS team’s and PENELLOPE team’s multi-band data initiatives (from ground and space) and the ULLYSES program—the latter providing an indispensable UV spectroscopic library of young high- and low-mass stars in the local universe. These comprehensive datasets serve as a critical foundation to maximize the scientific return of next-generation infrared flagships, including Euclid, Roman, and JWST. Operating in the X-ray regime, NewAthena will sustain and expand upon the legacy of earlier Athena mission designs, enabling the essential long-term monitoring of magnetic activity and high-energy environments in low-mass stars.

\section{The Substellar Connection: Euclid and Roman at the Low-Mass Frontier}

\noindent The wide-field surveys conducted by Euclid and the forthcoming Nancy Grace Roman Space Telescope are expected to identify thousands of young brown dwarfs, planetary-mass objects, and potentially free-floating planets in nearby star-forming regions such as Orion Molecular Cloud Complex and Perseus Molecular Cloud (\cite{2020AJ....160..123J}; \cite{2022A&A...664A.136B}; \cite{Zhang_2025}). Euclid’s deep near-infrared imaging and Roman’s combination of wide-field infrared sensitivity, high astrometric precision, and time-domain capability will enable the construction of unprecedented censuses of the low-mass population across diverse Galactic environments.

\noindent However, detecting these objects is only the first step; determining their physical origin remains one of the central open questions in star and planet formation theory. Ultraviolet spectroscopy provides one of the most powerful diagnostics for distinguishing between “star-like” formation via direct gravitational collapse and fragmentation of molecular clouds, and “planet-like” formation within circumstellar disks followed by dynamical ejection (\cite{Xuan_2024}). Measurements of UV excess emission, continuum veiling, and high-energy line profiles can reveal whether magnetospheric accretion processes persist smoothly into the planetary-mass regime 
(M<13 $M_{Jup}$; \cite{2021AJ....161..244Z}), although the exact deuterium-burning boundary is model-dependent and should not be interpreted as a strict physical division between planets and brown dwarfs (\cite{2011ApJ...727...57S}; \cite{2012A&A...547A.105M}).

\noindent The Hubble Space Telescope remains uniquely capable of detecting UV accretion signatures and chromospheric activity in these intrinsically faint substellar objects (\cite{2021AJ....161..244Z}). In particular, observations with COS and STIS can probe hot gas associated with accretion shocks and magnetic activity, extending accretion diagnostics into a regime inaccessible to most other facilities (\cite{Pittman_2025}). By combining HST UV spectroscopy with the large statistical samples identified by Euclid and Roman, it becomes possible to determine whether accretion rates, magnetic activity, and disk evolution scale continuously from low-mass stars into the free-floating planetary-mass domain (\cite{2024A&A...685A.118A}). These measurements are essential for constraining the low-mass end of the Initial Mass Function, testing competing formation pathways for substellar objects, and establishing whether free-floating planetary-mass bodies represent the low-mass extension of star formation or a dynamically processed population originating in protoplanetary disks (\cite{2010ARA&A..48..339B}).

\section{A Strategic Temporal Anchor for the Habitable Worlds Observatory}

The future success of the Habitable Worlds Observatory will depend not only on its direct characterization capabilities, but also on the long-term UV legacy established by current observatories. In particular, the Hubble Space Telescope provides an irreplaceable bridge between present-day studies of stellar activity and the future atmospheric characterization of potentially habitable exoplanets.

\subsection{Multi-Decadal UV Variability}

Planetary habitability is intrinsically time-dependent and is strongly shaped by the cumulative high-energy radiation environment generated by the host star. Ultraviolet and X-ray emission associated with stellar magnetic activity (including flares, energetic particle events, coronal mass ejection environments, and long-term magnetic cycles; \cite{DeWarf_2010};\cite{2025ApJ...985..100D}) can profoundly influence the thermal structure, chemistry, and long-term stability of planetary atmospheres (\cite{2003ApJ...598L.121L}; \cite{Linsky2025}). High-energy photons deposit energy in the upper atmosphere, driving ionization, dissociation of key molecules (such as H2O, CO2, CH4, and O2), and hydrodynamic escape processes that can remove volatile species over geological timescales (\cite{2003ARA&A..41..429K}; \cite{2015AsBio..15..119L}). These effects are especially significant for close-in terrestrial planets orbiting active K- and M-dwarf stars (\cite{2019AREPS..47...67O}), where sustained EUV and FUV irradiation may erode atmospheres, alter surface conditions, and potentially produce abiotic biosignature false positives through non-equilibrium photochemistry (\cite{2022AN....34310111I}). Young stars are particularly important in this context because they exhibit substantially elevated UV and X-ray luminosities, faster rotation rates, and more frequent flare activity relative to mature solar-type stars (\cite{2010ApJ...714..384R}), implying that the earliest stages of planetary atmospheric evolution may be dominated by stellar activity.

\noindent By leveraging the multi-decade ultraviolet archive of the Hubble Space Telescope, the astronomical community can establish long-term UV variability baselines for nearby young and solar-type stars across activity-cycle timescales that are inaccessible to most current missions. These archival datasets provide a uniquely valuable temporal record for quantifying both stochastic variability (e.g., flares) and secular evolution in stellar chromospheric and transition-region emission (\cite{2016ApJ...820...89F, 2018ApJ...867...71L}). Such long-baseline characterization is essential for interpreting future observations from the HWO (\cite{2025JATIS..11d2236D}) because atmospheric spectra obtained at a single observational epoch may not represent the long-term equilibrium state of an exoplanet atmosphere. Instead, observed spectral signatures could reflect transient stellar activity (\cite{2022AJ....163..147G}), temporally enhanced photochemistry, or short-term atmospheric responses to recent irradiation events (\cite{2018ApJ...867...71L, Barclay_2021}). Continuous UV monitoring therefore provides the temporal and physical context required to connect stellar magnetic evolution to atmospheric escape, climate stability, and the long-term retention of potentially habitable environments (\cite{2022AN....34310111I}). In this sense, the HST UV archive functions not only as a historical record of stellar activity, but also as a foundational calibration dataset for interpreting biosignature and habitability assessments in the HWO era.

\subsection{Sustaining the Scientific and Technical Pipeline}

Equally important, continued investment in UV observational programs preserves the scientific and technical expertise required for the next generation of space-based UV observatories. High-resolution UV spectroscopy demands specialized knowledge in detector calibration, background subtraction, line-spread-function characterization, radiative transfer modeling, and time-domain analysis: capabilities that cannot be rapidly regenerated after long observational gaps. Large-scale HST UV legacy surveys conducted during the telescope’s remaining operational lifetime would therefore serve a dual purpose: they would provide the foundational high-energy datasets needed to interpret future observations from HWO, James Webb Space Telescope, Nancy Grace Roman Space Telescope, and Euclid, while simultaneously sustaining an experienced community of UV observers, instrumentalists, and theorists through the HWO development era.

\vspace{0.1cm}
\noindent \textit{This paper therefore argues that the final decade of HST operations should prioritize coordinated UV legacy surveys, ensuring that the major infrared and exoplanet observatories of the 2030s are anchored by a robust high-energy astrophysical framework.}

\vspace{0.1cm}
\noindent \textbf{Context}: This paper was submitted to STScI as a response to "Building a Roadmap for HST Science into the 2030s" Community White Papers on May 22nd, 2026.

\noindent \textbf{Acknowledgment}
F.Z. Majidi acknowledges support from the ASI-INAF Agreement no. 2021-12-HH.0 and Addendum 2021-12-HH.1-2024 "Missione Solar-C EUVST-Supporto scientifico di Fase B/C/D."

\bibliographystyle{plain} 
\bibliography{references}

@article{Herczeg_2023,
doi = {10.3847/1538-4357/acf468},
url = {https://doi.org/10.3847/1538-4357/acf468},
year = {2023},
month = {oct},
publisher = {The American Astronomical Society},
volume = {956},
number = {2},
pages = {102},
author = {Herczeg, Gregory J. and Chen, Yuguang and Donati, Jean-Francois and Dupree, Andrea K. and Walter, Frederick M. and Hillenbrand, Lynne A. and Johns-Krull, Christopher M. and Manara, Carlo F. and Günther, Hans Moritz and Fang, Min and Schneider, P. Christian and Valenti, Jeff A. and Alencar, Silvia H. P. and Venuti, Laura and Alcalá, Juan Manuel and Frasca, Antonio and Arulanantham, Nicole and Linsky, Jeffrey L. and Bouvier, Jerome and Brickhouse, Nancy S. and Calvet, Nuria and Espaillat, Catherine C. and Campbell-White, Justyn and Carpenter, John M. and Chang, Seok-Jun and Cruz, Kelle L. and Dahm, S. E. and Eislöffel, Jochen and Edwards, Suzan and Fischer, William J. and Guo, Zhen and Henning, Thomas and Ji, Tao and Jose, Jessy and Kastner, Joel H. and Launhardt, Ralf and Principe, David A. and Robinson, Connor E. and Serna, Javier and Siwak, Michal and Sterzik, Michael F. and Takasao, Shinsuke},
title = {Twenty-five Years of Accretion onto the Classical T Tauri Star TW Hya},
journal = {The Astrophysical Journal}
}

@ARTICLE{2024A&A...689A.330T,
       author = {{Temmink}, Milou and {van Dishoeck}, Ewine F. and {Gasman}, Danny and {Grant}, Sierra L. and {Tabone}, Beno{\^\i}t and {G{\"u}del}, Manuel and {Henning}, Thomas and {Barrado}, David and {Caratti o Garatti}, Alessio and {Glauser}, Adrian M. and {Kamp}, Inga and {Arabhavi}, Aditya M. and {Jang}, Hyerin and {Kurtovic}, Nicolas and {Perotti}, Giulia and {Schwarz}, Kamber and {Vlasblom}, Marissa},
        title = "{MINDS: The DR Tau disk: II. Probing the hot and cold H$_{2}$O reservoirs in the JWST-MIRI spectrum}",
      journal = {\aap},
         year = 2024,
        month = sep,
       volume = {689},
          eid = {A330},
        pages = {A330},
          doi = {10.1051/0004-6361/202450355},
archivePrefix = {arXiv},
       eprint = {2407.05070},
 primaryClass = {astro-ph.EP},
       adsurl = {https://ui.adsabs.harvard.edu/abs/2024A&A...689A.330T}
}

@INPROCEEDINGS{2023MmSAI..94b.281C,
       author = {{Covino}, S. and {Cristiani}, S. and {Alcal{\'a}}, J.~M. and {Alencar}, S.~H.~P. and {Balashev}, S.~A. and {Barbuy}, B. and {Bastian}, N. and {Battino}, U. and {Bissell}, L. and {Bristow}, P. and {Calcines}, A. and {Calderone}, G. and {Cambianica}, P. and {Carini}, R. and {Carter}, B. and {Cassisi}, S. and {Castilho}, B.~V. and {Cescutti}, G. and {Christlieb}, N. and {Cirami}, R. and {Conzelmann}, R. and {Coretti}, I. and {Cooke}, R. and {Cremonese}, G. and {Cunha}, K. and {Cupani}, G. and {da Silva}, A.~R. and {D'Auria}, D. and {De Caprio}, V. and {De Cia}, A. and {Dekker}, H. and {D'Elia}, V. and {De Silva}, G. and {Diaz}, M. and {Di Marcantonio}, P. and {D'Odorico}, V. and {Ernandes}, H. and {Evans}, C. and {Fitzsimmons}, A. and {Franchini}, M. and {G{\"a}nsicke}, B. and {Genoni}, M. and {Giribaldi}, R.~E. and {Gneiding}, C. and {Grazian}, A. and {Hansen}, C.~J. and {Hopgood}, J. and {Izzo}, L. and {Kosmalski}, J. and {La Forgia}, F. and {La Penna}, P. and {Landoni}, M. and {Lazzarin}, M. and {Lunney}, D. and {Maciel}, W. and {Marcolino}, W. and {Marconi}, M. and {Migliorini}, A. and {Miller}, C. and {Modigliani}, A. and {Noterdaeme}, P. and {Oggioni}, L. and {Opitom}, C. and {Pariani}, G. and {Pilecki}, B. and {Piranomonte}, S. and {Quirrenbach}, A. and {Redaelli}, E.~M.~A. and {Pereira}, C.~B. and {Randich}, S. and {Rossi}, S. and {Sanchez-Janssen}, R. and {Schoeller}, M. and {Seifert}, W. and {Smiljanic}, R. and {Snodgrass}, C. and {Squalli}, O. and {Stilz}, I. and {St{\"u}rmer}, J. and {Trost}, A. and {Vanzella}, E. and {Ventura}, P. and {Verducci}, O. and {Waring}, C. and {Watson}, S. and {Wells}, M. and {Wright}, D. and {Zafar}, T. and {Zanutta}, A. and {Zins.}, G.},
        title = "{CUBES: a UV spectrograph for the future}",
    booktitle = {Memorie della Societa Astronomica Italiana},
         year = 2023,
       volume = {94},
        month = sep,
        pages = {281},
          doi = {10.36116/MEMSAIT_94N2.2023.281},
archivePrefix = {arXiv},
       eprint = {2212.12791},
 primaryClass = {astro-ph.IM},
       adsurl = {https://ui.adsabs.harvard.edu/abs/2023MmSAI..94b.281C}
}

@ARTICLE{2012ApJ...756...24S,
       author = {{Scholz}, Alexander and {Jayawardhana}, Ray and {Muzic}, Koraljka and {Geers}, Vincent and {Tamura}, Motohide and {Tanaka}, Ichi},
        title = "{Substellar Objects in Nearby Young Clusters (SONYC). VI. The Planetary-mass Domain of NGC 1333}",
      journal = {\apj},
         year = 2012,
        month = sep,
       volume = {756},
       number = {1},
          eid = {24},
        pages = {24},
          doi = {10.1088/0004-637X/756/1/24},
archivePrefix = {arXiv},
       eprint = {1207.1449},
 primaryClass = {astro-ph.SR},
       adsurl = {https://ui.adsabs.harvard.edu/abs/2012ApJ...756...24S}
}

@article{Muzic_2014,
doi = {10.1088/0004-637X/785/2/159},
url = {https://doi.org/10.1088/0004-637X/785/2/159},
year = {2014},
month = {apr},
publisher = {The American Astronomical Society},
volume = {785},
number = {2},
pages = {159},
author = {Mužić, Koraljka and Scholz, Alexander and Geers, Vincent C. and Jayawardhana, Ray and Martí, Belén López},
title = {SUBSTELLAR OBJECTS IN NEARBY YOUNG CLUSTERS (SONYC). VIII. SUBSTELLAR POPULATION IN LUPUS 3*},
journal = {The Astrophysical Journal}
}

@ARTICLE{2012ApJ...744...60G,
       author = {{Green}, James C. and {Froning}, Cynthia S. and {Osterman}, Steve and {Ebbets}, Dennis and {Heap}, Sara H. and {Leitherer}, Claus and {Linsky}, Jeffrey L. and {Savage}, Blair D. and {Sembach}, Kenneth and {Shull}, J. Michael and {Siegmund}, Oswald H.~W. and {Snow}, Theodore P. and {Spencer}, John and {Stern}, S. Alan and {Stocke}, John and {Welsh}, Barry and {B{\'e}land}, St{\'e}phane and {Burgh}, Eric B. and {Danforth}, Charles and {France}, Kevin and {Keeney}, Brian and {McPhate}, Jason and {Penton}, Steven V. and {Andrews}, John and {Brownsberger}, Kenneth and {Morse}, Jon and {Wilkinson}, Erik},
        title = "{The Cosmic Origins Spectrograph}",
      journal = {\apj},
         year = 2012,
        month = jan,
       volume = {744},
       number = {1},
          eid = {60},
        pages = {60},
          doi = {10.1088/0004-637X/744/1/6010.1086/141956},
archivePrefix = {arXiv},
       eprint = {1110.0462},
 primaryClass = {astro-ph.IM},
       adsurl = {https://ui.adsabs.harvard.edu/abs/2012ApJ...744...60G}
}

@INPROCEEDINGS{2023ASPC..534..567P,
       author = {{Pascucci}, I. and {Cabrit}, S. and {Edwards}, S. and {Gorti}, U. and {Gressel}, O. and {Suzuki}, T.~K.},
        title = "{The Role of Disk Winds in the Evolution and Dispersal of Protoplanetary Disks}",
    booktitle = {Protostars and Planets VII},
         year = 2023,
       editor = {{Inutsuka}, S. and {Aikawa}, Y. and {Muto}, T. and {Tomida}, K. and {Tamura}, M.},
       series = {Astronomical Society of the Pacific Conference Series},
       volume = {534},
        month = jul,
        pages = {567},
          doi = {10.48550/arXiv.2203.10068},
archivePrefix = {arXiv},
       eprint = {2203.10068},
 primaryClass = {astro-ph.EP},
       adsurl = {https://ui.adsabs.harvard.edu/abs/2023ASPC..534..567P}
}

@article{Espaillat_2022,
doi = {10.3847/1538-3881/ac479d},
url = {https://doi.org/10.3847/1538-3881/ac479d},
year = {2022},
month = {feb},
publisher = {The American Astronomical Society},
volume = {163},
number = {3},
pages = {114},
author = {Espaillat, C. C. and Herczeg, G. J. and Thanathibodee, T. and Pittman, C. and Calvet, N. and Arulanantham, N. and France, K. and Serna, Javier and Hernández, J. and Kóspál, Á. and Walter, F. M. and Frasca, A. and Fischer, W. J. and Johns-Krull, C. M. and Schneider, P. C. and Robinson, C. and Edwards, Suzan and Ábrahám, P. and Fang, Min and Erkal, J. and Manara, C. F. and Alcalá, J. M. and Alecian, E. and Alexander, R. D. and Alonso-Santiago, J. and Antoniucci, Simone and Ardila, David R. and Banzatti, Andrea and Benisty, M. and Bergin, Edwin A. and Biazzo, Katia and Briceño, César and Campbell-White, Justyn and Cleeves, L. Ilsedore and Coffey, Deirdre and Eislöffel, Jochen and Facchini, Stefano and Fedele, D. and Fiorellino, Eleonora and Froebrich, Dirk and Gangi, Manuele and Giannini, Teresa and Grankin, K. and Günther, Hans Moritz and Guo, Zhen and Hartmann, Lee and Hillenbrand, Lynne A. and Hinton, P. C. and Kastner, Joel H. and Koen, Chris and Maucó, K. and Mendigutía, I. and Nisini, B. and Panwar, Neelam and Principe, D. A. and Robberto, Massimo and Sicilia-Aguilar, A. and Valenti, Jeff A. and Wendeborn, J. and Williams, Jonathan P. and Xu, Ziyan and Yadav, R. K.},
title = {The ODYSSEUS Survey. Motivation and First Results: Accretion, Ejection, and Disk Irradiation of CVSO 109},
journal = {The Astronomical Journal}
}

@article{Ardila_2013,
doi = {10.1088/0067-0049/207/1/1},
url = {https://doi.org/10.1088/0067-0049/207/1/1},
year = {2013},
month = {jun},
publisher = {The American Astronomical Society},
volume = {207},
number = {1},
pages = {1},
author = {Ardila, David R. and Herczeg, Gregory J. and Gregory, Scott G. and Ingleby, Laura and France, Kevin and Brown, Alexander and Edwards, Suzan and Johns-Krull, Christopher and Linsky, Jeffrey L. and Yang, Hao and Valenti, Jeff A. and Abgrall, Hervé and Alexander, Richard D. and Bergin, Edwin and Bethell, Thomas and Brown, Joanna M. and Calvet, Nuria and Espaillat, Catherine and Hillenbrand, Lynne A. and Hussain, Gaitee and Roueff, Evelyne and Schindhelm, Rebecca N. and Walter, Frederick M.},
title = {HOT GAS LINES IN T TAURI STARS},
journal = {The Astrophysical Journal Supplement Series}
}

@ARTICLE{2017RSOS....470114E,
       author = {{Ercolano}, Barbara and {Pascucci}, Ilaria},
        title = "{The dispersal of planet-forming discs: theory confronts observations}",
      journal = {Royal Society Open Science},
         year = 2017,
        month = apr,
       volume = {4},
       number = {4},
          eid = {170114},
        pages = {170114},
          doi = {10.1098/rsos.170114},
archivePrefix = {arXiv},
       eprint = {1704.00214},
 primaryClass = {astro-ph.EP},
       adsurl = {https://ui.adsabs.harvard.edu/abs/2017RSOS....470114E}
}

@article{Zhang_2025,
doi = {10.3847/1538-3881/addfcb},
url = {https://doi.org/10.3847/1538-3881/addfcb},
year = {2025},
month = {jul},
publisher = {The American Astronomical Society},
volume = {170},
number = {2},
pages = {64},
author = {Zhang, Zhoujian and Mollière, Paul and Fortney, Jonathan J. and Marley, Mark S.},
title = {ELemental Abundances of Planets and Brown Dwarfs Imaged around Stars (ELPIS). II. The Jupiter-like Inhomogeneous Atmosphere of the First Directly Imaged Planetary-mass Companion 2MASS 1207 b},
journal = {The Astronomical Journal}
}

@ARTICLE{2020AJ....160..123J,
       author = {{Johnson}, Samson A. and {Penny}, Matthew and {Gaudi}, B. Scott and {Kerins}, Eamonn and {Rattenbury}, Nicholas J. and {Robin}, Annie C. and {Calchi Novati}, Sebastiano and {Henderson}, Calen B.},
        title = "{Predictions of the Nancy Grace Roman Space Telescope Galactic Exoplanet Survey. II. Free-floating Planet Detection Rates}",
      journal = {\aj},
         year = 2020,
        month = sep,
       volume = {160},
       number = {3},
          eid = {123},
        pages = {123},
          doi = {10.3847/1538-3881/aba75b},
archivePrefix = {arXiv},
       eprint = {2006.10760},
 primaryClass = {astro-ph.EP},
       adsurl = {https://ui.adsabs.harvard.edu/abs/2020AJ....160..123J}
}

@ARTICLE{2022A&A...664A.136B,
       author = {{Bachelet}, E. and {Specht}, D. and {Penny}, M. and {Hundertmark}, M. and {Awiphan}, S. and {Beaulieu}, J.-P. and {Dominik}, M. and {Kerins}, E. and {Maoz}, D. and {Meade}, E. and {Nucita}, A.~A. and {Poleski}, R. and {Ranc}, C. and {Rhodes}, J. and {Robin}, A.~C.},
        title = "{Euclid-Roman joint microlensing survey: Early mass measurement, free floating planets, and exomoons}",
      journal = {\aap},
         year = 2022,
        month = aug,
       volume = {664},
          eid = {A136},
        pages = {A136},
          doi = {10.1051/0004-6361/202140351},
archivePrefix = {arXiv},
       eprint = {2202.09475},
 primaryClass = {astro-ph.EP},
       adsurl = {https://ui.adsabs.harvard.edu/abs/2022A&A...664A.136B}
}

@BOOK{1998apsf.book.....H,
       author = {{Hartmann}, Lee},
        title = "{Accretion Processes in Star Formation}",
         year = 1998,
       volume = {32},
       adsurl = {https://ui.adsabs.harvard.edu/abs/1998apsf.book.....H}
}

@ARTICLE{2014A&A...561A...2A,
       author = {{Alcal{\'a}}, J.~M. and {Natta}, A. and {Manara}, C.~F. and {Spezzi}, L. and {Stelzer}, B. and {Frasca}, A. and {Biazzo}, K. and {Covino}, E. and {Randich}, S. and {Rigliaco}, E. and {Testi}, L. and {Comer{\'o}n}, F. and {Cupani}, G. and {D'Elia}, V.},
        title = "{X-shooter spectroscopy of young stellar objects. IV. Accretion in low-mass stars and substellar objects in Lupus}",
      journal = {\aap},
         year = 2014,
        month = jan,
       volume = {561},
          eid = {A2},
        pages = {A2},
          doi = {10.1051/0004-6361/201322254},
archivePrefix = {arXiv},
       eprint = {1310.2069},
 primaryClass = {astro-ph.SR},
       adsurl = {https://ui.adsabs.harvard.edu/abs/2014A&A...561A...2A}
}

@ARTICLE{2017A&A...600A..20A,
       author = {{Alcal{\'a}}, J.~M. and {Manara}, C.~F. and {Natta}, A. and {Frasca}, A. and {Testi}, L. and {Nisini}, B. and {Stelzer}, B. and {Williams}, J.~P. and {Antoniucci}, S. and {Biazzo}, K. and {Covino}, E. and {Esposito}, M. and {Getman}, F. and {Rigliaco}, E.},
        title = "{X-shooter spectroscopy of young stellar objects in Lupus. Accretion properties of class II and transitional objects}",
      journal = {\aap},
         year = 2017,
        month = apr,
       volume = {600},
          eid = {A20},
        pages = {A20},
          doi = {10.1051/0004-6361/201629929},
archivePrefix = {arXiv},
       eprint = {1612.07054},
 primaryClass = {astro-ph.SR},
       adsurl = {https://ui.adsabs.harvard.edu/abs/2017A&A...600A..20A}
}

@ARTICLE{2017A&A...604A.127M,
       author = {{Manara}, C.~F. and {Testi}, L. and {Herczeg}, G.~J. and {Pascucci}, I. and {Alcal{\'a}}, J.~M. and {Natta}, A. and {Antoniucci}, S. and {Fedele}, D. and {Mulders}, G.~D. and {Henning}, T. and {Mohanty}, S. and {Prusti}, T. and {Rigliaco}, E.},
        title = "{X-shooter study of accretion in Chamaeleon I. II. A steeper increase of accretion with stellar mass for very low-mass stars?}",
      journal = {\aap},
         year = 2017,
        month = aug,
       volume = {604},
          eid = {A127},
        pages = {A127},
          doi = {10.1051/0004-6361/201630147},
archivePrefix = {arXiv},
       eprint = {1704.02842},
 primaryClass = {astro-ph.SR},
       adsurl = {https://ui.adsabs.harvard.edu/abs/2017A&A...604A.127M}
}

@INPROCEEDINGS{2023ASPC..534..539M,
       author = {{Manara}, C.~F. and {Ansdell}, M. and {Rosotti}, G.~P. and {Hughes}, A.~M. and {Armitage}, P.~J. and {Lodato}, G. and {Williams}, J.~P.},
        title = "{Demographics of Young Stars and their Protoplanetary Disks: Lessons Learned on Disk Evolution and its Connection to Planet Formation}",
    booktitle = {Protostars and Planets VII},
         year = 2023,
       editor = {{Inutsuka}, S. and {Aikawa}, Y. and {Muto}, T. and {Tomida}, K. and {Tamura}, M.},
       series = {Astronomical Society of the Pacific Conference Series},
       volume = {534},
        month = jul,
        pages = {539},
          doi = {10.48550/arXiv.2203.09930},
archivePrefix = {arXiv},
       eprint = {2203.09930},
 primaryClass = {astro-ph.SR},
       adsurl = {https://ui.adsabs.harvard.edu/abs/2023ASPC..534..539M}
}

@article{Carvalho_2024,
doi = {10.3847/2041-8213/ad74eb},
url = {https://doi.org/10.3847/2041-8213/ad74eb},
year = {2024},
month = {sep},
publisher = {The American Astronomical Society},
volume = {973},
number = {2},
pages = {L40},
author = {Carvalho, Adolfo S. and Hillenbrand, Lynne A. and France, Kevin and Herczeg, Gregory J.},
title = {A Far-ultraviolet-detected Accretion Shock at the Star–Disk Boundary of FU Ori},
journal = {The Astrophysical Journal Letters}
}

@ARTICLE{2026arXiv260519710D,
       author = {{Das}, Gautam and {Hillenbrand}, Lynne A. and {Carvalho}, Adolfo S.},
        title = "{A Parameterized YSO Accretion Disk Model with Increasing Accretion Rate: Predicted Outburst Lightcurves}",
      journal = {arXiv e-prints},
         year = 2026,
        month = may,
          eid = {arXiv:2605.19710},
        pages = {arXiv:2605.19710},
          doi = {10.48550/arXiv.2605.19710},
archivePrefix = {arXiv},
       eprint = {2605.19710},
 primaryClass = {astro-ph.SR},
       adsurl = {https://ui.adsabs.harvard.edu/abs/2026arXiv260519710D}
}

@ARTICLE{2021A&A...652A..72A,
       author = {{Alcal{\'a}}, J.~M. and {Gangi}, M. and {Biazzo}, K. and {Antoniucci}, S. and {Frasca}, A. and {Giannini}, T. and {Munari}, U. and {Nisini}, B. and {Harutyunyan}, A. and {Manara}, C.~F. and {Vitali}, F.},
        title = "{GIARPS High-resolution Observations of T Tauri stars (GHOsT). III. A pilot study of stellar and accretion properties}",
      journal = {\aap},
         year = 2021,
        month = aug,
       volume = {652},
          eid = {A72},
        pages = {A72},
          doi = {10.1051/0004-6361/202140918},
archivePrefix = {arXiv},
       eprint = {2106.10724},
 primaryClass = {astro-ph.SR},
       adsurl = {https://ui.adsabs.harvard.edu/abs/2021A&A...652A..72A}
}

@ARTICLE{2011ApJ...743..105I,
       author = {{Ingleby}, Laura and {Calvet}, Nuria and {Bergin}, Edwin and {Herczeg}, Gregory and {Brown}, Alexander and {Alexander}, Richard and {Edwards}, Suzan and {Espaillat}, Catherine and {France}, Kevin and {Gregory}, Scott G. and {Hillenbrand}, Lynne and {Roueff}, Evelyne and {Valenti}, Jeff and {Walter}, Frederick and {Johns-Krull}, Christopher and {Brown}, Joanna and {Linsky}, Jeffrey and {McClure}, Melissa and {Ardila}, David and {Abgrall}, Herv{\'e} and {Bethell}, Thomas and {Hussain}, Gaitee and {Yang}, Hao},
        title = "{Near-ultraviolet Excess in Slowly Accreting T Tauri Stars: Limits Imposed by Chromospheric Emission}",
      journal = {\apj},
         year = 2011,
        month = dec,
       volume = {743},
       number = {2},
          eid = {105},
        pages = {105},
          doi = {10.1088/0004-637X/743/2/105},
archivePrefix = {arXiv},
       eprint = {1110.6312},
 primaryClass = {astro-ph.SR},
       adsurl = {https://ui.adsabs.harvard.edu/abs/2011ApJ...743..105I}
}

@article{Ingleby_2013,
doi = {10.1088/0004-637X/767/2/112},
url = {https://doi.org/10.1088/0004-637X/767/2/112},
year = {2013},
month = {apr},
publisher = {The American Astronomical Society},
volume = {767},
number = {2},
pages = {112},
author = {Ingleby, Laura and Calvet, Nuria and Herczeg, Gregory and Blaty, Alex and Walter, Frederick and Ardila, David and Alexander, Richard and Edwards, Suzan and Espaillat, Catherine and Gregory, Scott G. and Hillenbrand, Lynne and Brown, Alexander},
title = {ACCRETION RATES FOR T TAURI STARS USING NEARLY SIMULTANEOUS ULTRAVIOLET AND OPTICAL SPECTRA},
journal = {The Astrophysical Journal}
}

@INPROCEEDINGS{2014prpl.conf..475A,
       author = {{Alexander}, R. and {Pascucci}, I. and {Andrews}, S. and {Armitage}, P. and {Cieza}, L.},
        title = "{The Dispersal of Protoplanetary Disks}",
    booktitle = {Protostars and Planets VI},
         year = 2014,
       editor = {{Beuther}, Henrik and {Klessen}, Ralf S. and {Dullemond}, Cornelis P. and {Henning}, Thomas},
        month = jan,
        pages = {475-496},
          doi = {10.2458/azu_uapress_9780816531240-ch021},
archivePrefix = {arXiv},
       eprint = {1311.1819},
 primaryClass = {astro-ph.EP},
       adsurl = {https://ui.adsabs.harvard.edu/abs/2014prpl.conf..475A}
}

@ARTICLE{2001ApJ...553L.153H,
       author = {{Haisch}, Jr., Karl E. and {Lada}, Elizabeth A. and {Lada}, Charles J.},
        title = "{Disk Frequencies and Lifetimes in Young Clusters}",
      journal = {\apjl},
         year = 2001,
        month = jun,
       volume = {553},
       number = {2},
        pages = {L153-L156},
          doi = {10.1086/320685},
archivePrefix = {arXiv},
       eprint = {astro-ph/0104347},
 primaryClass = {astro-ph},
       adsurl = {https://ui.adsabs.harvard.edu/abs/2001ApJ...553L.153H}
}

@ARTICLE{2018MNRAS.477.5191R,
       author = {{Richert}, A.~J.~W. and {Getman}, K.~V. and {Feigelson}, E.~D. and {Kuhn}, M.~A. and {Broos}, P.~S. and {Povich}, M.~S. and {Bate}, M.~R. and {Garmire}, G.~P.},
        title = "{Circumstellar disc lifetimes in numerous galactic young stellar clusters}",
      journal = {\mnras},
         year = 2018,
        month = jul,
       volume = {477},
       number = {4},
        pages = {5191-5206},
          doi = {10.1093/mnras/sty949},
archivePrefix = {arXiv},
       eprint = {1804.05076},
 primaryClass = {astro-ph.SR},
       adsurl = {https://ui.adsabs.harvard.edu/abs/2018MNRAS.477.5191R}
}

@article{Gorti_2009,
doi = {10.1088/0004-637X/705/2/1237},
url = {https://doi.org/10.1088/0004-637X/705/2/1237},
year = {2009},
month = {oct},
publisher = {The American Astronomical Society},
volume = {705},
number = {2},
pages = {1237},
author = {Gorti, U. and Dullemond, C. P. and Hollenbach, D.},
title = {TIME EVOLUTION OF VISCOUS CIRCUMSTELLAR DISKS DUE TO PHOTOEVAPORATION BY FAR-ULTRAVIOLET, EXTREME-ULTRAVIOLET, AND X-RAY RADIATION FROM THE CENTRAL STAR},
journal = {The Astrophysical Journal}
}

@article{OBERG20211,
title = {Astrochemistry and compositions of planetary systems},
journal = {Physics Reports},
volume = {893},
pages = {1-48},
year = {2021},
note = {Astrochemistry and compositions of planetary systems},
issn = {0370-1573},
doi = {https://doi.org/10.1016/j.physrep.2020.09.004},
url = {https://www.sciencedirect.com/science/article/pii/S0370157320303446},
author = {Karin I. Öberg and Edwin A. Bergin}
}

\end{document}